\documentclass[10pt]{iopart}
\usepackage{graphicx}
\usepackage{dcolumn}
\usepackage{bm}
\usepackage{mathrsfs}
\usepackage{amssymb}
\usepackage[breaklinks=true,colorlinks=true,linkcolor=blue,urlcolor=blue,citecolor=blue]{hyperref}
\begin{document}

\title {
How random are random numbers generated using photons?}
\author{Aldo Solis, Al\'\i {~}M. Angulo Mart\'\i nez, Roberto Ram\'\i rez Alarc\'on, Hector Cruz Ram\'\i rez, Alfred B. U'Ren, Jorge G. Hirsch}

\address{$^1$Instituto de Ciencias Nucleares, Universidad Nacional
Aut\'onoma de M\'exico, Apdo. Postal 70-543, M\'exico 04510 D.F.}

\ead{hirsch@nucleares.unam.mx}

\begin{abstract}
Randomness is fundamental in quantum theory, with many philosophical and practical implications.
In this paper we discuss the concept of algorithmic randomness, which provides a quantitative method
to assess the Borel normality of a given sequence of numbers, a necessary condition for it to be considered random. 
We use Borel normality as a tool to investigate the randomness of ten sequences of bits generated from the differences 
between detection times of photon pairs generated by spontaneous parametric downconversion. These sequences are shown to fulfil the randomness criteria without difficulties.
As deviations from Borel normality  for photon-generated random number sequences have been reported in previous work,
a strategy to understand these diverging findings is outlined.
\end{abstract}

\vspace{2pc}

\ioptwocol

\section{Introduction}

For many decades after Quantum Mechanics was formally established, it was relatively easy
to live with the fact that it only allows the prediction of probabilities of certain results of experiments.
The description and manipulation of condensed matter, molecules, atoms, atomic nuclei, 
and subnuclear particles was indeed very successful, and the statistical nature of the results,
describing a huge number of similar processes, as in scattering and decay, fitted quite well with this probabilistic interpretation.

In the last few decades, the ability to manipulate individual quantum objects (e.g. molecules, atoms, and photons) and even to place many of
them in a single quantum state, as in a Bose-Einstein condensate, has emerged.  The experiments along these lines which are possible nowadays invite one to
ponder on the quantum mechanical description of individual systems, and to engineer them to obtain 
technologically useful devices.

The emerging view is, for us, both surprising and challenging.
Nearly ten years ago, Anton Zeilinger wrote: ``The discovery that individual events are irreducibly random is probably one of the
most significant findings of the twentieth century. ... for the individual event in quantum
physics, not only do we not know the cause, there is no cause. The instant when a
radioactive atom decays, or the path taken by a photon behind a half-silvered beamsplitter
are objectively random "\cite{art:Zei05}.

This provocative statement helps to visualize the relevance that randomness has in our description of the
physical world. According to this view, reality and information are two sides of the same coin. Randomness,
complementarity and entanglement emerge from the fact that from individual measurements there is a finite amount
of information available.  It is postulated that an elementary system can only give a definite result in one specific
measurement. Other independent measurements must then be irreducibly random \cite{Zei99}.

The consequences of these assumptions are both philosophical and practical. Random numbers are widely employed for 
classical computation in science and industry. Monte Carlo and other numerical methods require the use of random numbers,
which are demanded to be both efficiently generated and having proved randomness \cite{Kat10}. 

When the intrinsic quality of quantum randomness is accepted as a postulate, a world of applications emerges. 
Quantum key distribution is a cryptographic process whose security is guaranteed by the quantum randomness \cite{Niel02}. A cryptographically secure random number generator that does not require any assumption about the internal working of the device has been proposed \cite{Sca10,Pir10}. Such a strong form of randomness generation is impossible classically and possible in quantum systems only if certified by a Bell inequality violation.
``Private randomness" is defined by the presence of correlations that cannot be reproduced with local variables. It is quantified by the violation of Bell inequalities, and is associated with the impossibility to predict a given string employing a classical computer and classical information \cite{Sca10,Pir10,chris13,bancal14,nieto14}. 
Quantum contextuality has also been invoked to certify randomness in a random number generator \cite{Um13}.

Is the randomness of quantum phenomena a physical assumption that is testable? Many attempts have been made to answer this question.
Recently a comprehensive suite of tests, developed at the National Institute of Standards and Technology (NIST) to assess the quality of computer-based random number generators \cite{art:nist}, was employed to study the randomness of single-photon polarization measurement outcomes, using pairs of photons generated by spontaneous parametric downconversion (SPDC). No statistically significant deviations from randomness were observed \cite{Bra10}.

Has quantum randomness been experimentally proved? The above-mentioned results suggest that a quantum-generated random sequence looks as random as a computer-generated one. But there are deep differences between these sources of random numbers. Quantum randomness can be
proven incomputable; that is, it is not exactly reproducible by any algorithm, while software-generated random numbers, known as pseudo-random, can be reproduced if the computer code and the seed are known. Is it possible to distinguish between them?  Calude et al. \cite{art:Calude:ExpEvi} performed finite tests of randomness inspired by algorithmic information theory, analyzing algorithmic randomness, the strongest possible form of incomputability. They performed tests of randomness on pseudo-random strings (finite sequences) generated with software (Mathematica, Maple), which are cyclic (so, strongly computable), the bits of $\pi$, which are computable, but not cyclic, and strings produced by quantum
measurements (with the commercial device Quantis and by the Vienna IQOQI group). They report that all tests produced evidence -with different degrees of statistical significance- of differences between quantum and non-quantum sources. 

\begin{figure}
\centering
\includegraphics[width=1.0\columnwidth]{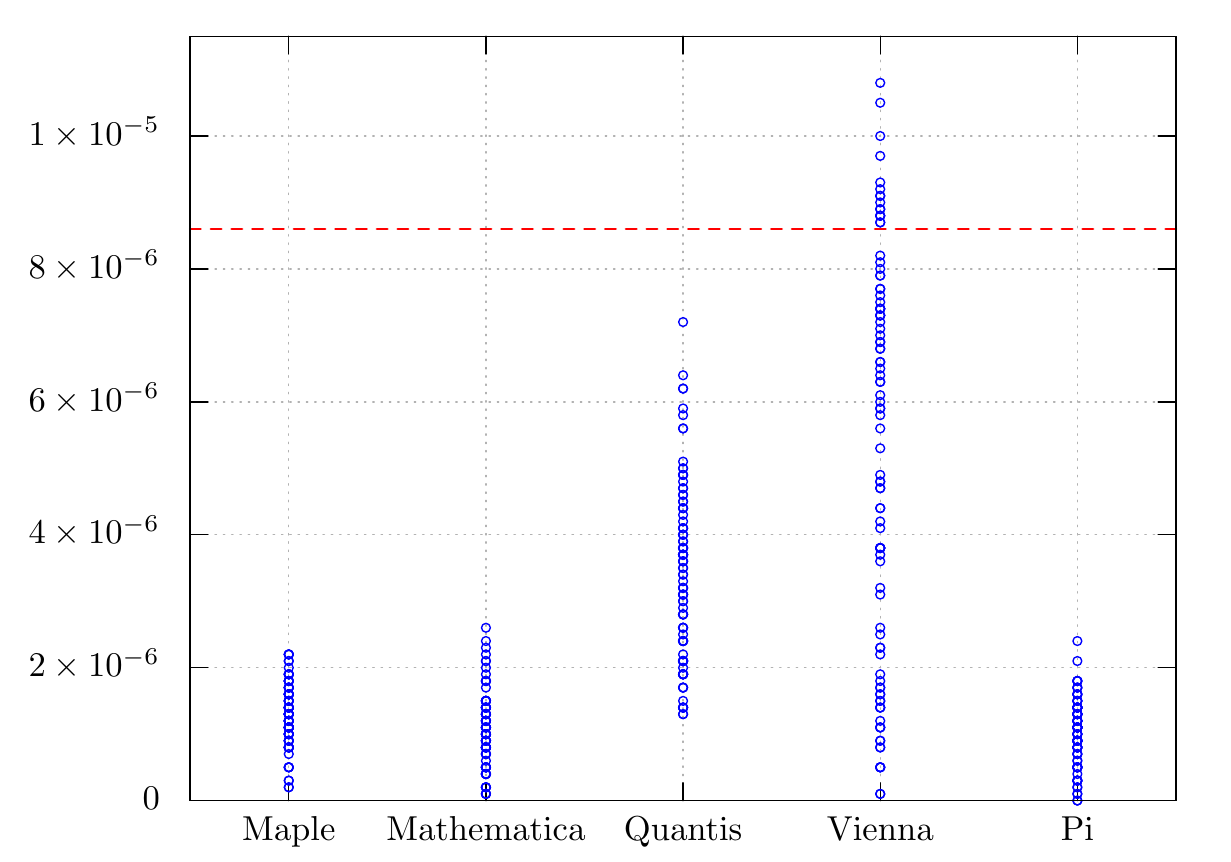}
\caption{Maximal relative deviations from Borel normality reported in  
\cite{art:Calude:ExpEvi}. The red dashed line represents the maximum possible 
deviation allowed by Borel normality.  The blue circles represent the (absolute value of the) deviations from
$1/2^m$, for different orders $m$ (see Section 3, below).  Some sequences produced by the Vienna
group are not Borel normal.}
\label{calude}
\end{figure} 

Figure \ref{calude} displays the maximal relative deviations from Borel normality reported in \cite{art:Calude:ExpEvi}. It shocked us to observe that the non-computable, photon-generated random sequences depart from Borel normality far more than the pseudo random numbers obtained from Mathematica, Maple and the digits of $\pi$. In particular, the sequences produced by the Vienna group, obtained from an attenuated laser impinging on a beamsplitter, cannot be validated as random under this criterion. 

Puzzled by these results, we have generated random number employing the detection times of photon pairs generated by SPDC, and analyzed them with the tools of algorithmic randomness. We have found that, at variance with the findings of \cite{art:Calude:ExpEvi},  they pass all the randomness tests with flying colors. These results call for a more detailed analysis, comparing different sources of single photons and different ways to generate random bits from their detection times; this constitutes work in progress in our group.

In what follows we present a short review of the challenges faced in defining randomness, the algorithmic information approach, and the Borel normality test,  which combined with the algorithmic complexity approach provides a necessary but insufficient test of randomness. We describe in some detail the experimental setup, the procedure to generate the bit sequence from the photon arrival times, and the analysis. We close with some conclusions and open questions.

\section{Randomness}

Randomness  in physics, is related with two main ideas: first with the lack of information about a system, for instance 
every time a coin is tossed we ignore the initial conditions of the event so that we 
cannot predict the result, and second, with the intrinsically unpredictable behavior of a quantum system, as in  
a photon impinging on a beam splitter. 
In both cases we use a probabilistic approach to describe phenomena regardless
of their conceptual differences and sometimes we combine both in 
the density matrix, but there are situations where we  want to talk about randomness itself, i.e. characterize and/or quantify 
randomness. We therefore need a definition of randomness and a theoretical framework that allows us to develop and use these ideas.

There have been some works trying to define randomness through different mathematical
objects. For example the concept of normal number due to Borel \cite{art:borel}
formalizes the notion of a random real number. One successful attempt that has 
become a mathematical theory is Algorithmic Information theory, also known as
Kolmogorov complexity. It is based on the idea of patterns in a mathematical 
object.

To motivate the definition of randomness in algorithmic information theory,
consider a system having a property with the two possible outcomes, 0 and 1, 
with equal probabilities of measuring each of them.

If we measure it repeatedly the result may look like one of these cases:
\begin{eqnarray*}
0101010101010101010101010101010101\cdots \\
0100011011000001010011100101110111\cdots \\ 
0100010110100101111010101001001011\cdots \\
\label{predecible}
\end{eqnarray*}

These sets are noticeably different.  
The first one does not look random at all, because it has a strong pattern: after
a 0 the next result will be a 1. We can sum up this set of data with the recipe  ``repeat '01' indefinitely''. 
The second case looks random but actually has a pattern. It was generated writing 0 and 1, then the combinations 00 01 10 11, then
the combinations with three symbols 000 001 \ldots etc., and is known as the Champernowne's constant.
It serves as a cautionary example of the difficulty in identifying a random sequence, which 
can look random despite being defined by a  very simple pattern. 
How can we determine if the third one has a pattern?

It is possible to formalize the intuitive relationship between randomness and the lack of patterns in a sequence in a
definition like this:
\begin{quote}
``A sequence without patterns is random''
\end{quote}

Although this looks like a very simple definition, we need to define what a
pattern is. In the previous examples we described a pattern in such a way that anyone can reproduce the sequence just by following the steps, so everyone could use this description, in principle, in the same way that a computer executes a program. 

Suppose that we have a program that generates a given sequence.  Does that mean that the sequence has a pattern? 
As a useful illustration, the program 

$$PRINT ``0100010110100101111010101001001011''$$ 

\noindent evidently can reproduce the third sequence and in general the same technique can be 
used for every sequence simply by placing the actual sequence as part of  the code. However, this implies that
the program is longer than the sequence itself and it does not make much sense considering it as a pattern; 
so we define a pattern as a program whose output is a given sequence in such  a way that this program is shorter than the sequence.

The considerations above lead us to a simple definition:
\begin{quote}
There are no patterns in a random sequence, i.e. every program that 
outputs a random sequence is longer than the sequence itself.
\end{quote}

This definition isn't fully formalized.  The notion of program needs to 
be written in terms of universal Turing machines and we need to specify the 
symbols allowed in the code.   A full treatment of the concept of  algorithmic randomness is beyond
the scope of this informal exposition.  The reader interested in a complete description is invited to read \cite{book:Calude} and \cite{book:Li}. 

Unfortunately this definition holds some surprises for us. Suppose that we wish to
determine whether or not a sequence is random;  this is indeed a very natural question to ask.
In principle, we would need to run all programs with a shorter length than the length of the 
sequence.  If any of these programs were to give us the desired sequence as the output, an underlying pattern would then have been determined to exist, and we would conclude that the sequence is not random. But we don't know in advance if a given program will halt or not. There is a close relationship between algorithmic randomness, the halting problem and G\"odel's incompleteness \cite{calude05}. 

The halting problem \cite{art:turing} in a computer can be summarized as follows: suppose that we have an algorithm designed to determine whether a given program will halt, then we could build a new algorithm that halts if the program does not halt and does not halt if the program does. What will happen if we feed this new program with its own code? It will halt if, and only if it doesn't halt, which is an evident paradox. Therefore there is no algorithm capable of deciding whether a sequence is random or not.

\section{Borel Normality}
We may be disappointed because algorithmic randomness is difficult to apply in a real case, but there are other ways to approach this definition of randomness.

In the sequences above the ``probability'' of getting 0 or 1 is equal;  this is the case for the sequence  $010101\cdots$.
What about the probability of getting $01$, following a subdivision of the string into symbols comprised of two digits? In this case, this
probability is one and the probabilities of getting $00$,$10$ and $11$ are in all three cases zero. 
In contrast, we expect a random sequence to lead to equal probabilities for all these cases:

\begin{eqnarray*}
P(0)=P(1)=\frac{1}{2} \\
P(00)=P(01)=P(10)=P(11)=\frac{1}{4}  \\
\vdots 
\end{eqnarray*}

We can generalize this restriction on the probabilities as:
\begin{equation}
P(< m \mbox{ bits\ sequence}>)=\frac{1}{2^{m}}.
\end{equation}

A set of numbers satisfying this property is called Borel normal. 
It is closely related with the Normality concept in 
real numbers developed by Borel \cite{art:borel}. 
Naturally, there must be a restriction on $m$ because in a sequence of 4 
symbols we cannot try sequences longer than 4 symbols. 

We can analyze finite sequences that may not fulfil exactly the 
conditions above but are close to doing so, for example: 
$$P(00)=0.251$$ 

This can be expressed mathematically as
 \cite{book:Calude}
\begin{equation*}
\left| P(00)-\frac{1}{2^{2}} \right| < \epsilon
\label{condicion}
\end{equation*}
where $\epsilon$ is a ``small'' number.

Although there is no algorithm which can determine the randomness of sequences,
it is possible to relate algorithmic randomness with the parameters $m$ and $
\epsilon$. 
In \cite{art:calude-borel}  it is shown that almost all algorithmic random 
strings are Borel normal, satisfying 
\begin{equation} 
\left| P(< m \mbox{ bits sequence}>)-\frac{1}{2^{m}} \right| < \sqrt{\frac{log \,n}{n}}
\label{cond2}
\end{equation}
where $n$ is the length of the complete sequence and 
\begin{equation}
m\le log\ log\ n  .
\label{mcondicion}
\end{equation}

We will refer to this condition as Borel Normality. This is not a sufficient condition for randomness but allows 
us to discard some sequences as clearly not random. Its advantage is that checking this 
condition is an algorithmic procedure which can be applied to any sequence.

In \cite{art:Calude:ExpEvi} this condition is applied to binary sequences 
obtained from a quantum experiment, based on an attenuated laser beam impinging on a beam splitter. The authors found that these sequences ``were outside the expected range for $m = 3$  and $m = 4$''. 
In this work we analyze a related experimental setup, 
where we employ 
the differences in arrival times of SPDC photon pairs as a means to generate random bits.

\section{Experimental setup}

Our experimental work is based on the process of spontaneous parametric downconversion (SPDC) in which a laser pump beam illuminates a crystal with a $\chi^{(2)}$ nonlinearity, leading to 
the annihilation of pump photons and the emission of photon pairs, typically referred to as signal and idler~\cite{burnham70}.    In the case of a continuous-wave pump at frequency $\omega_p$, signal and idler photons are spectrally anti-correlated so that if a signal photon is detected with frequency $\omega$, the conjugate idler must have a frequency $\omega_p-\omega$.  Likewise, in the idealized situation of a plane-wave pump (which we may approximate through a Gaussian beam with large beam radius at the beamwaist), photon pairs are anti-correlated in transverse wavevector, i.e. if a signal photon is detected with transverse wavevector $\textbf{k}^\bot$, the conjugate idler photon  must have transverse wavevector $-\textbf{k}^\bot$.  

The quantum state of the emitted photon pairs can be written as $|\Psi\rangle=|\mbox{vac}\rangle+\eta |\Psi_2\rangle$ in terms of the vacuum $|\mbox{vac}\rangle$, the two-photon component $|\Psi_2\rangle$, and of a constant $\eta$ related to the conversion efficiency.   Under the assumptions a continuous-wave, plane-wave pump $ |\Psi_2\rangle$ may be expressed as~\cite{vicent10}

\begin{equation}\label{E:state}
 |\Psi_2\rangle=\int d \omega \int d \textbf{k}^\bot F(\omega,\textbf{k}^\bot) |\omega,\textbf{k}^\bot  \rangle_s |\omega_p-\omega,-\textbf{k}^\bot  \rangle_i,
\end{equation}

\noindent written in terms of a joint amplitude function $F(\omega,\textbf{k}^\bot)$, and  where $|\omega,\textbf{k}^\bot  \rangle_\mu$ represents a single-photon Fock state with frequency $\omega$ and transverse wavevector $\textbf{k}^\bot$ for mode $\mu$, with $\mu=s,i$ for the signal ($s$) and idler ($i$).    In writing the two-photon state, we have assumed that the parametric downconversion process is in the  spontaneous regime, so that the appearance of multiple-pair events can be neglected.   
This assumption is valid if the parametric gain is sufficiently low; experimentally, we restrict the pump power so that the process remains spontaneous.  In all likelihood, a similar experiment and analysis carried out in the high-gain, stimulated regime would yield different results from those presented on this paper.  

Note that the state in Eq.~\ref{E:state} is entangled since it cannot be factored into a direct product of separate states $|S\rangle $  (for the signal photon) and $|I\rangle$ (for the idler photon) as $|\Psi\rangle=|S\rangle |I\rangle$.  While many experimental works have focused on the presence of quantum entanglement in photon pairs, in this paper we exploit another important aspect of SPDC photon pairs: they are emitted at random times.    

\begin{figure}[h]
\centering
\includegraphics[width=1.0\columnwidth]{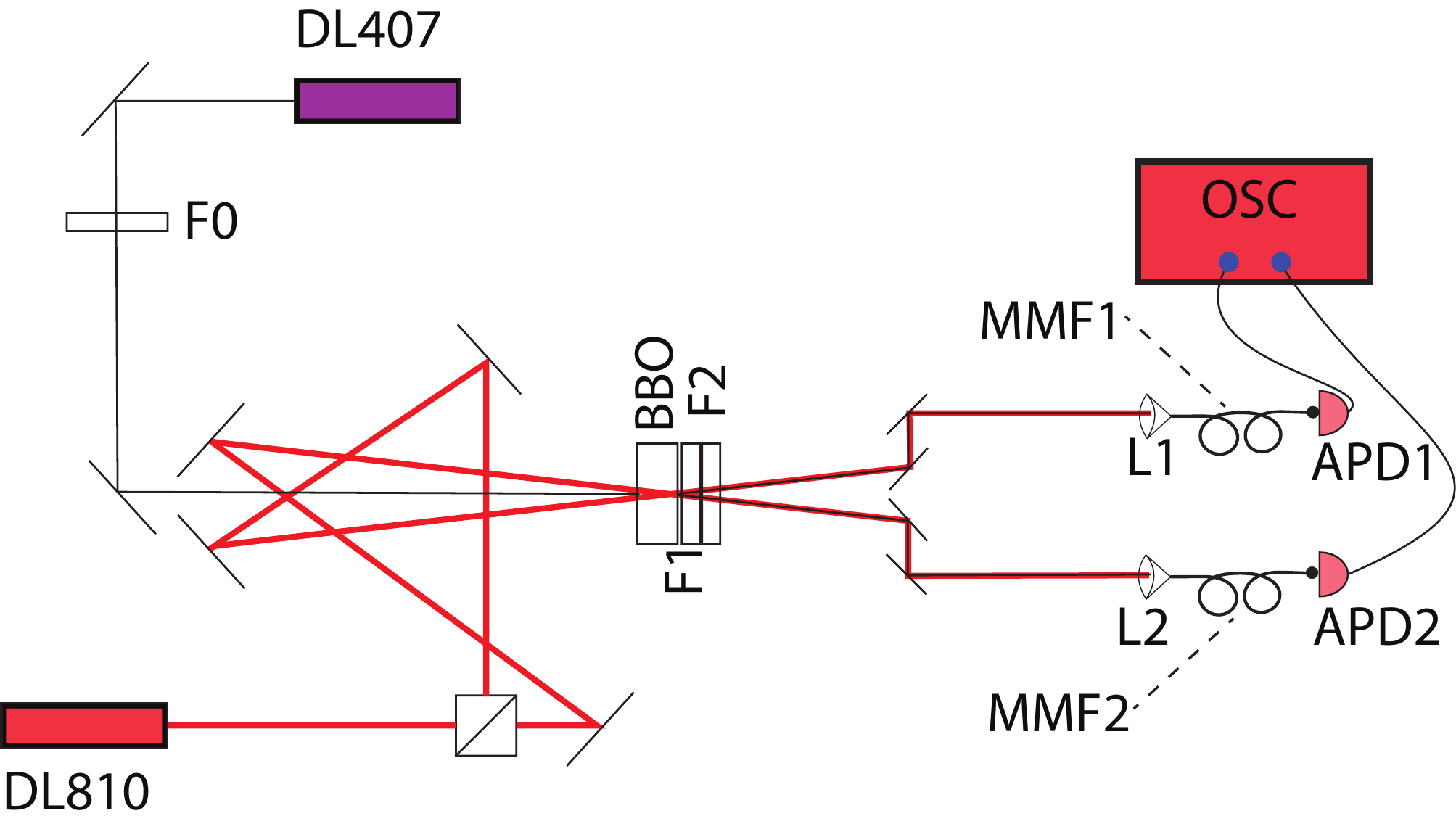}
\caption{Experimental setup used for obtaining random sequences from a SPDC photon-pair source}
\label{montaje}
\end{figure}

Our experimental setup is shown in Fig.~\ref{montaje}.  We have used as pump a beam from a diode laser (DL407) centered at $407$nm with $\sim60$mW power, and as nonlinear medium a  $\beta$ barium borate (BBO) crystal of $1$mm length.  A Schott BG-39 coloured glass filter (F0) is used to remove non-ultraviolet photons from the pump beam.   The BBO crystal, which is negative uniaxial,  was cut so that the angle subtended by the optic axis with respect pump beam axis is $\theta_{\mbox{pm}}=29.2^\circ$ which yields phasematching for the generation of frequency-degenerate, non-collinear photon pairs.  Signal and idler photons are emitted on diametrically opposed portions of an emission cone centred on the pump beam axis, in our case with a $3.6^\circ$ half opening angle.  Pump photons are suppressed by transmitting the signal and idler modes through a long-pass filter which transmits wavelengths  $\lambda>488$nm (F1), followed by a bandpass filter centred at $800$nm with a $40$nm bandwidth (F2).

Note that in order to set up and correctly align the fiber collection modes, the signal and idler paths are initially simulated using a separate diode laser centred at $810$nm (DL810).  The beam from this laser is split into two branches, and each one is reflected with a mirror so as to meet on the crystal's centre plane in such a way that the paths of these two branches emerge from the crystal in the directions expected for the emitted signal and idler photons.  Alignment of the collection lenses and fibers is significantly easier with these classical beams than with SPDC light.

Collection of the signal and idler photons can be carried out on any two diametrically-opposed locations on the emission ring.  Each of the signal and idler collection modes is defined by an aspheric lens with $f=8$mm focal length (L1 and L2) which focuses incoming light into the core of a multi-mode fiber with a $50\mu$m diameter (MMF1 and MMF2).  The plane defined by the two collection fibers is chosen for convenience to be parallel to the optical table.    

Each of the two photon-collection fibers leads to a silicon-based avalanche photodiode (APD1 and APD2), which emits an electronic pulse for each detection event, discriminated on its rising edge resulting in a $4$ns-long standard NIM pulse.  Because we are interested in the time series which results from the detection times, we connect the two detector outputs to a $2.5$GHz digital oscilloscope (OSC).  We program the oscilloscope to subdivide a detection span of $0.512$s into $256 \times 10^6$ time bins, so that each bin has a duration of $2$ns.    The voltage from the APD output is recorded at each time bin, for  each of the signal and idler channels, thus obtaining two separate time series composed of voltage values.      These times series are post-processed, so that those bins with a voltage  $V$ which satisfies  $|V|>V_{th}$ with a threshold value of $V_{th}=450$mV are assigned a value of $1$, while those bins for which $|V|<V_{th}$ are assigned a value of $0$.   This leads to two separate (for the signal and idler) strings $s_n$ and $i_n$ of $256\times10^6$ digits, each with values $0$ or $1$.      Likewise, for each pair of signal and idler time series, we generate a third time series defined as $c_n=s_n \times i_n$, which corresponds to those bins for which there are coincident detection events in the two channels.    We have observed, by averaging over several hundred experimental runs, that the average number of single-channel detection events during the detection span of $0.512$s  is around $8.5 \times 10^{5}$, while the average number of coincidence detection events is around $8 \times 10^4$.

\section{Obtaining bits from data}

Obtaining random bits from data has a considerable complexity.
The binary strings in the
previous section have a bias because, most of the time bins will have value of ``0'' corresponding to no detected photon.  In other words, 
there are many more zeroes than ones. 

In order to obtain a sequence that looks more
random we first compute the time difference between subsequent detection events. This is equivalent to counting
the number of zeroes between ones, to which we can clearly assign an integer number.
These time intervals are described by an exponential distribution which can be 
understood in mathematical terms like this:
let $\lambda$ be the probability per unit time of observing a detection at any instant of 
time, then the probability of not observing a detection in a short time interval 
 $\delta t$ is $1-\lambda\delta t$; if we consider a finite time interval $t$
and divide it into $n$ intervals with duration $\frac{t}{n}$ where $n$ is large, 
the probability of not having a detection in $t$ is equal to the product of the 
probabilities for each short time interval,  then:
$$ (1-\frac{\lambda t}{n})(1-\frac{\lambda t}{n})\ldots =(1-\frac{\lambda t}{n})^{n}.$$
Note that the right hand side of the previous expression becomes $e^{-\lambda t}$ as $n$ tends to infinity. 
It is relevant to remark that the above result relies on the absence of correlations between detections at different times.

\begin{figure}
\centering
\includegraphics[width=1.0\columnwidth]{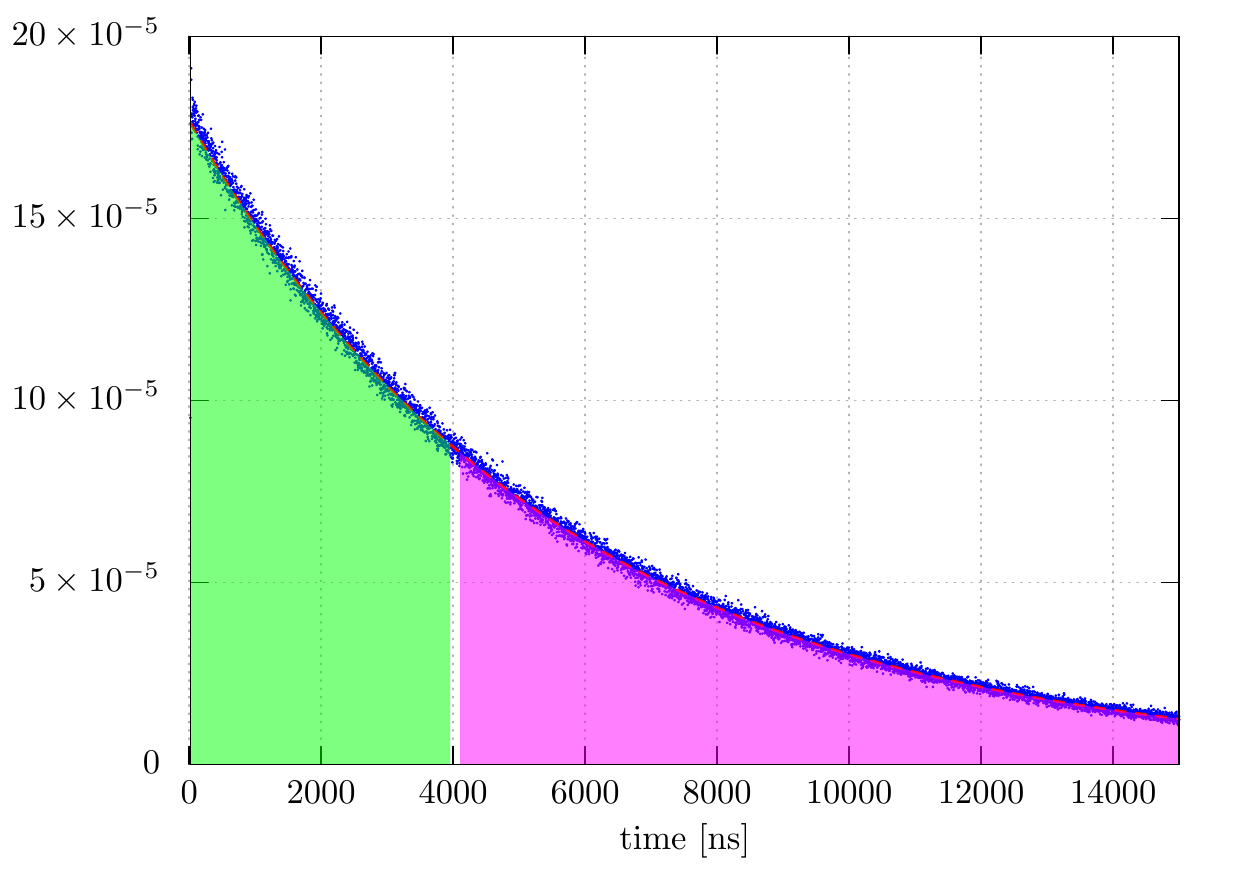}
\caption{Distribution of time differences between subsequent detection events. 
Each event in the green zone yields a $0$ value, while those in the purple zone yield a $1$ value.}
\label{distribucion}
\end{figure}

Figure~\ref{distribucion} shows an example of an experimentally-measured time interval distribution for our SPDC photon pairs detected in coincidence,  along with a fit to an exponential distribution. 
Note that our APD's have a dead time of around $T_d=20$ns, so that following a detection event in any of the two channels, the detector is unable to register further events during a time interval of $T_d$ duration.   Of course, this will impact the time interval distribution for short times.   We have based our analysis below on a truncated time interval distribution, so as to exclude the above features which appear at short times smaller than $t_0 = 2 \,T_d$. As the time intervals follow an exponential distribution, removing the short time intervals is equivalent to redefining them as $t \rightarrow t - t_0$. 
Note that while the interval distribution reaches times greater than $15\mu$s, with a mean time of around $4\mu$s, the omission of the first $40$ns of this temporal range is expected to have only a minor effect on our procedure for obtaining random sequences.   

Thus, using our knowledge about the distribution, random bits can be obtained, by dividing the possible time values into two bins:
those lower than $x$ and those greater than $x$, where $x$ is the mean time, such that

$$\int^{x}_{0}\lambda \, e^{-\lambda t}dt = \int^{\infty}_{x} \lambda \, e^{-\lambda t}dt=\frac{1}{2} .$$

Each individual time difference $t$ obtained from the experiment allows us to generate a random
bit, $0$ if the $t<x$ (green zone in figure \ref{distribucion}) and $1$ if $t>x$ (purple zone). 
 The fraction of 0's and 1's in each string departs from the exact $1/2$ value due to the fluctuations around the average exponential distribution. These deviations from $1/2$ are quantified by the standard deviation shown in the $m=1$ row of Table 1.  Using this method, we generated $10$ sequences of 
$10^6$ bits using the string of coincident detection events $c_n$, obtained from the quantum source described in the experimental setup.
Note that this method could be improved upon by dividing the possible time values
into more bins i.e. $4$,$8$,$16$, so we can obtain more bits per detection; see for 
example~\cite{art:tru_ran_num_gen,art:pho_arr_tim_qua,art:pra_fas_qua_ran}.

\section{Results}

We have experimentally generated ten sequences, which have an equal count of zeroes and ones with a maximum
discrepancy of $0.1$\%,  after a careful adjustment of the mean difference time $x$.
From condition~\ref{mcondicion} we have
\begin{eqnarray}
m &< &log\ log\ 10^6 = 4.3.
\end{eqnarray}
Therefore, the maximum possible value of $m$ in our case is $4$.

\begin{figure}
\centering
\includegraphics[width=1.0\columnwidth]{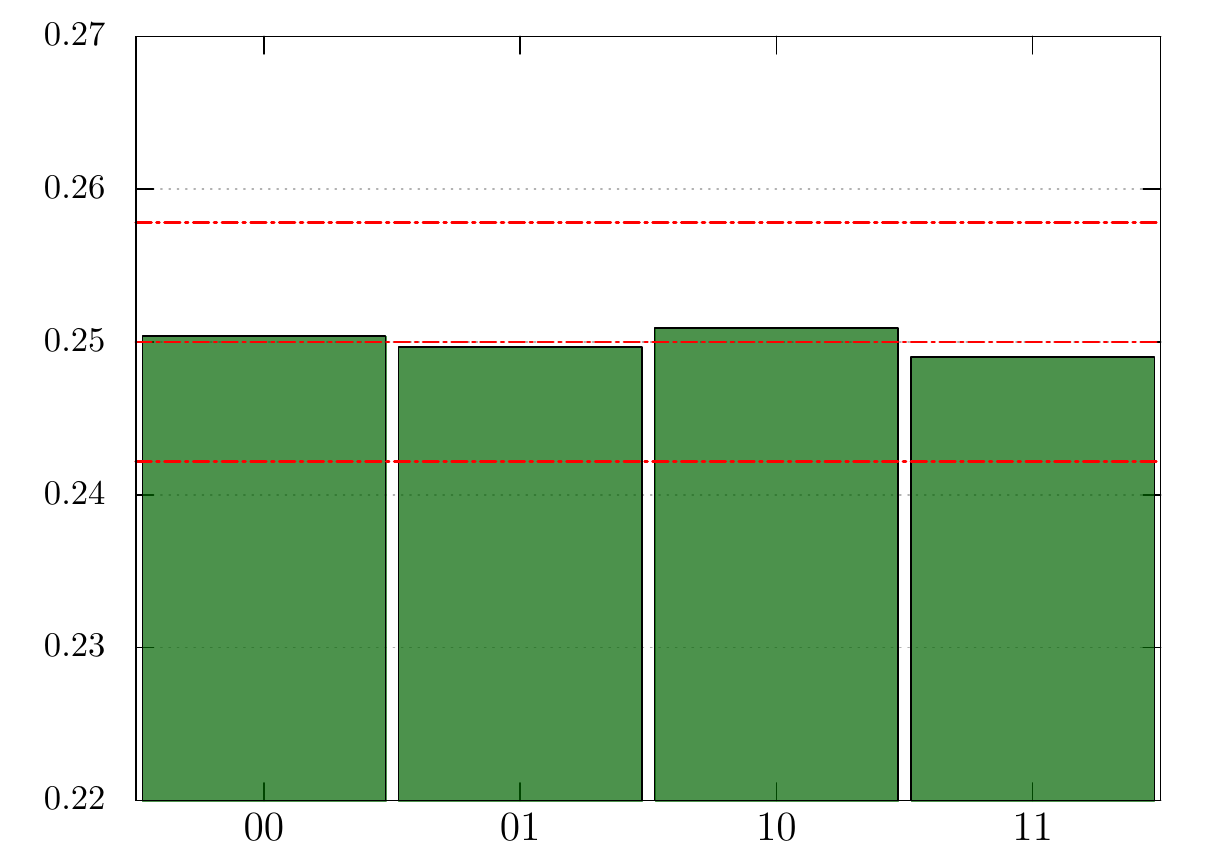}
\caption{Borel normality analysis for $m=2$, using sequence $2$. The top and bottom red lines are the 
maximum deviations allowed by Borel normality.}
\label{columnas2}
\end{figure}

\begin{figure}
\centering
\includegraphics[width=1.0\columnwidth]{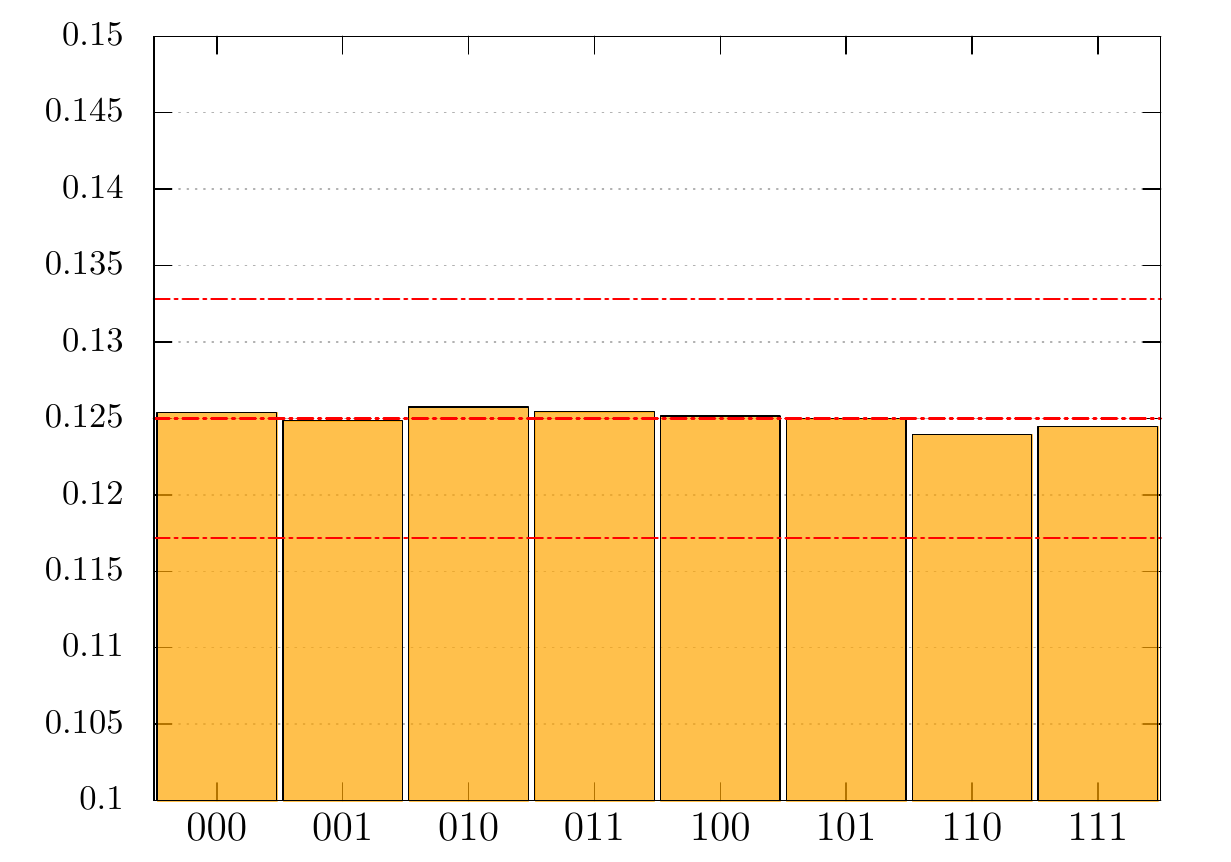}
\caption{Borel normality analysis for $m=3$, using sequence $2$. The top and bottom red lines are the 
maximum deviations allowed by Borel normality. }
\label{columnas3}
\end{figure}

\begin{figure}
\centering
\includegraphics[width=1.0\columnwidth]{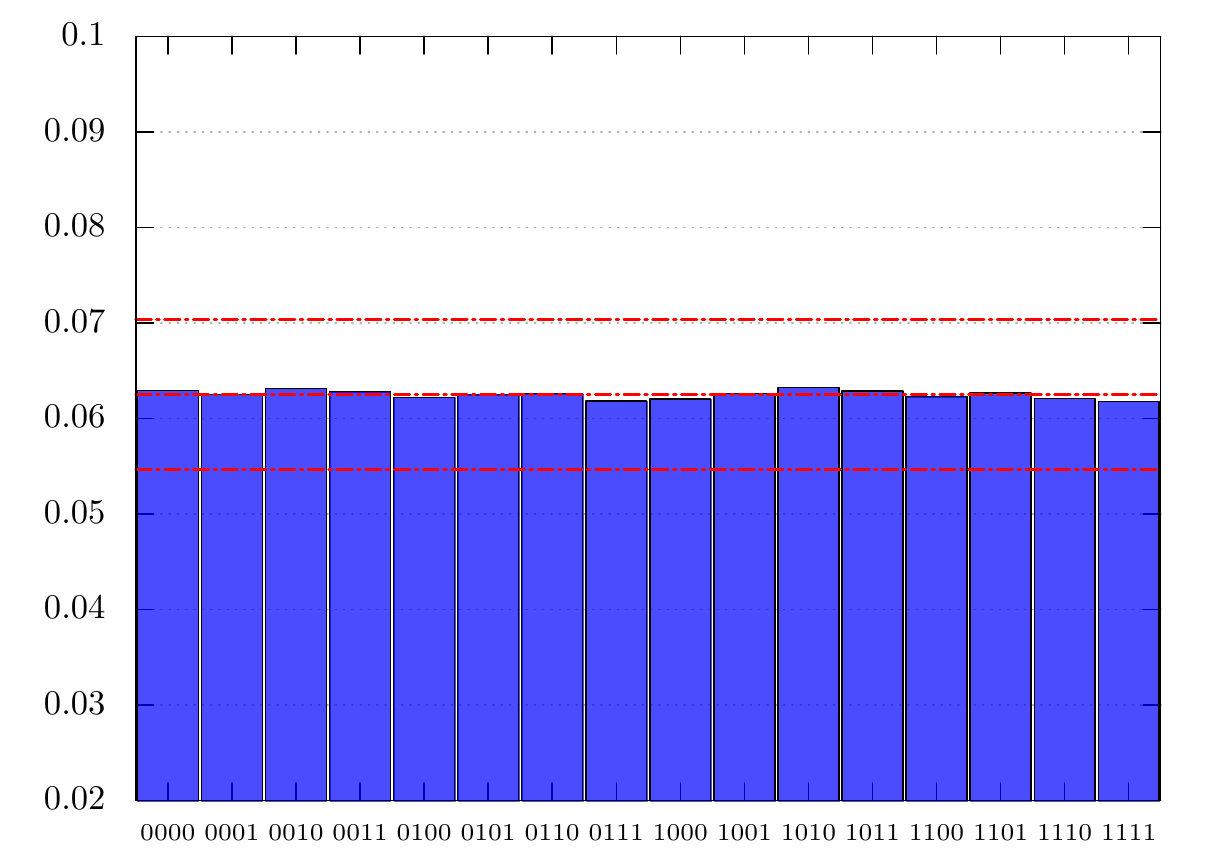}
\caption{Borel normality analysis for $m=4$, using sequence $2$. The top and bottom red lines are the 
maximum deviations allowed by Borel normality.}
\label{columnas4}
\end{figure}

For a given $m$, there are $2^m$ different strings of length forming the set $S_m$. Thus, $S_2=\{ 00, 01, 10, 11 \}$,   $S_3=\{ 000, 001, 010, 100, 011, 101, 110,$ $111 \}$ and so on. We evaluate the probability of occurrence of a given string $i \in S_m$ as $P(i) = N(i) / N_m$, where
 $N(i)$ is the number of times that the string $i$ is present in a sequence subdivided into segments of $m$ symbols,  and $N_m = Int[n/m]$ is the total  number of strings of length $m$ in the sequence of length $n$. For instance, in the sequence $01010101$ we find that $N(01)=4$ and $N(00)=N(10)=N(11)=0$.  Note that we define the strings $i$ sequentially in such a way that no two strings overlap; for the specific example above this means that the string ``10'' does not appear.

In figures \ref{columnas2},\ref{columnas3},\ref{columnas4} we present the results 
of the analysis for one 
sequence. Each histogram shows $P(i)$. It  is represented in a different range, because the mean 
value is $0.25$ in \ref{columnas2}, $0.125$ in \ref{columnas3} and $0.0625$ in \ref{columnas4}. The top and bottom dashed red lines represent the maximum and minimum possible value allowed by equation \ref{cond2} respectively, the central red line corresponds to the mean value expected. It can be seen that all cases easily satisfy the Borel normality condition
$$ \left| P(i)-\frac{1}{2^{m}} \right| < \sqrt{\frac{log 10^6}{10^6}} = 0.00441 .$$

Because we are mainly interested in the deviations from the mean value for 
every string, it makes sense to look at the standard deviation ($\sigma$) of the 
probabilities.
The standard deviation is defined as

\begin{equation}
\sigma^2_m = \frac {1}{2^m} \sum_{i \in S_m} \left(P(i) - \frac{1}{2^{m}} \right)^2 \label{Eq:SD}
\end{equation}

It follows from Eq. \ref{cond2} that, for the sequence to be considered Borel normal, all $\sigma_m$ values must satisfy the condition

\begin{equation} 
\sigma_m  < \frac{log\, n}{n} = 0.00441.
\end{equation} 

In table \ref{tabla}  we display the values of the standard deviations of four sequences. All of them  
fulfill the above condition for $m=1,2,3,4$, being about an order of magnitude smaller that the limit imposed
by algorithmic randomness on Borel normality.

\begin{table}
\begin{centering}
\begin{tabular}{||c|c|c|c|c||}
	\hline
	\hline
	\multicolumn{5}{||c||}{standard deviation $\sigma_m$} \\
	\hline
$m$	& seq.1 & seq. 2 & seq. 3 & seq. 4 \\
\hline
1 & 0.000980 & 0.000405 & 0.000542 & 0.000760 \\
\hline
2 & 0.001314 & 0.000750 & 0.000494& 0.000564 \\
\hline
3 & 0.000633& 0.000312 & 0.000535 & 0.000492 \\
\hline
4 & 0.000619 & 0.000421 & 0.000460 & 0.000451 \\
\hline
\hline
\end{tabular}
\end{centering}
\caption{Standard deviation of $P(i)$ values from $1/2^m$ (see Eq.~\ref{Eq:SD}), for different values of $m$, computed for four different sequences.}
\label{tabla}
\end{table}

\begin{figure}
\centering
\includegraphics[width=1.0\columnwidth]{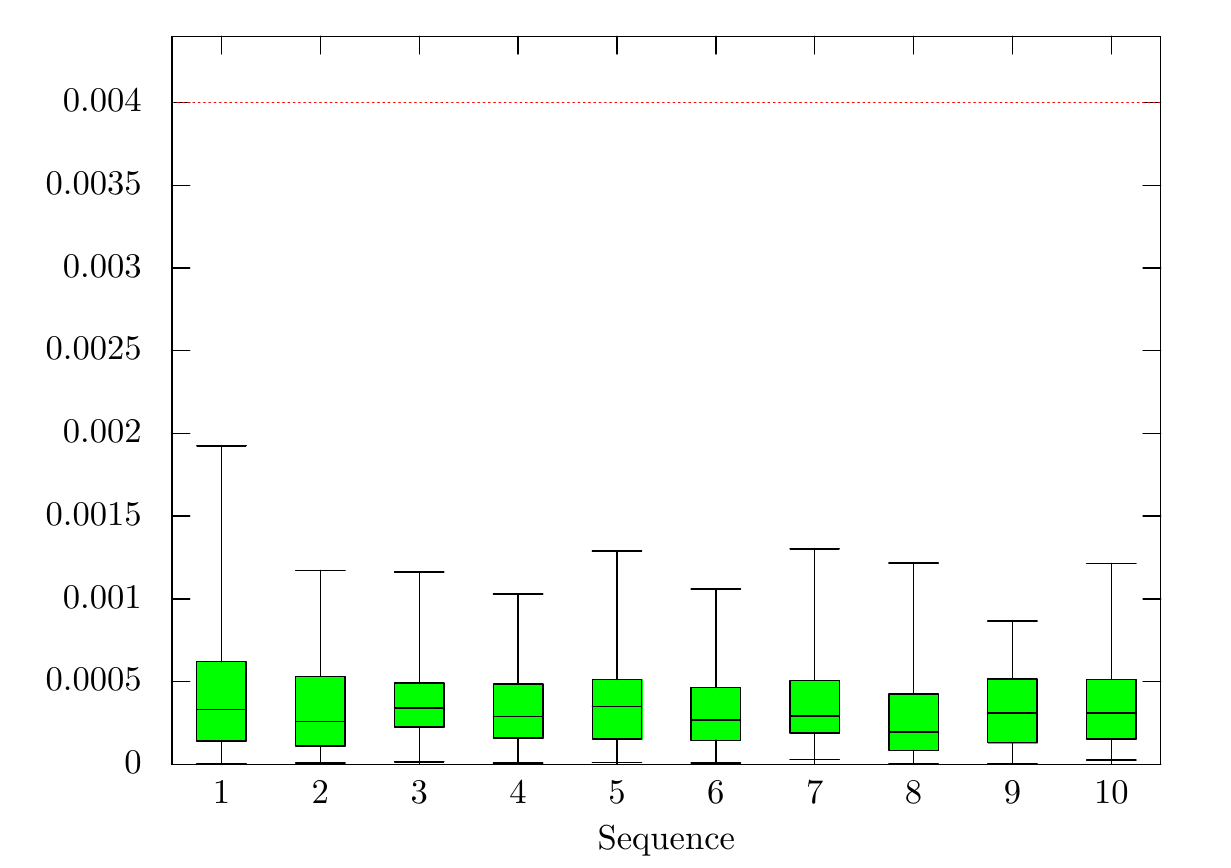}
\caption{Deviations from the mean value for each sequence. The red line represents the maximum deviation allowed by Borel normality.
}
\label{boxplot}
\end{figure}

In figure \ref{boxplot} we present, for each of the ten sequences, their deviations
from the expected mean value. The box-and-whisker plots, inspired in the ones used in \cite{art:Calude:ExpEvi}, display
in short horizontal lines, from bottom to top, the minimum value, first quantile, median, third quantile and 
maximum value, of the difference $| P(i)-\frac{1}{2^{m}} |$, including the results for $m=$2, 3 and 4. The whisker and box plot allows us to see the large difference 
between the maximum deviations from Borel normality and the value imposed by condition \ref{cond2}. The maximal deviations from the mean value reach up to 34\% of the limit set by condition \ref{cond2} for $m=1$, 47\% for $m=2$, 40\% for $m=3$ and 27\% for $m=4$.

Returning to the motivation for the present investigation, we were puzzled by the results reported in  \cite{art:Calude:ExpEvi}, where the sequences of pseudo-random numbers passed without difficulty the Borel normality test, while the sequences of random numbers built employing the photon detections produced by the Vienna group failed, having  maximal deviations from the mean value reach up to 27\% of the limit set by condition \ref{cond2} for $m=1$, 127\% for $m=2$, 103\% for $m=3$ and 105\% for $m=4$.

We have based our analysis on the coincident-event sequences $c_n$.  Note that since SPDC photons are born in pairs, ideally the single-channel sequences $s_n$ and $i_n$ would be identical to $c_n$,   Realistic experiments are affected by optical losses and by spurious detection events from noise sources and from dark counts.  We did in fact perform the Borel normality analysis on the single-channel sequences with the result that sequences $s_n$ and $i_n$ do satisfy the bounds for Borel normality imposed by Eq. \ref{cond2}, albeit with larger deviations from the expected fractions ($81\%$ for $m=2$, $54\%$ for $m=3$, and $37\%$ for $m=4$) as compared to the sequence $c_n$.

In  \cite{art:Calude:ExpEvi} the random bits were obtained with a different experimental setup.
The signals of the Vienna 
group were generated with photons from a weak blue LED light source which impinged on
a non-polarising beamsplitter with two output ports associated with the
values `0' and `1', respectively. There was no pre- or post-processing of the raw data
stream, however the output was constantly monitored. The signals of the QUANTIS device
are produced in a similar way, but due to hardware imbalances which are difficult to overcome at this level, QUANTIS processes
these raw data by unbiasing the sequence by a von Neumann-type normalization.
The sequences employed had $2^{32}\approx 4 \times 10^9$ bits.



Finally, we have used the NIST statistical test suite \cite{art:nist} to check the (intuitive or statistical) randomness of our generated bits.
Even though these tests are not directly related to algorithmic randomness,
it is worthwhile comparing these results with those of the Borel normality test. 
So as to have at least 100 sequences on which to run the NIST tests, we divided each string into ten sub-strings of equal length, ending up with $100$ strings of $10^5$ bits each.
Each test returns a P-value that must be greater than $0.01$ (in our case) to pass the test;
this value was calculated using the 100 sequences of random bits and at least 97 of the sequences need to pass
the test individually.  As expected, our sequences pass each of the tests in the suite; the results are presented in table \ref{tabla-nist}.

\begin{table}
\begin{centering}
\begin{tabular}{||c||c|c||}
	\hline
	\hline
	Test & P-value & Pass \\
	\hline
	Frequency & 0.191687 & 100/100 \\
	\hline
	Block Frequency & 0.021999 & 100/100 \\
	\hline
	Cumulative Sums & 0.171867 & 100/100 \\
	\hline
	Cumulative Sums & 0.319084 & 100/100 \\
	\hline
	Runs & 0.383827 & 98/100 \\ 
	\hline
	Longest Run & 0.224821 & 100/100 \\
	\hline 
	Rank & 0.019188 & 99/100 \\
	\hline 
	FFT & 0.867692 & 100/100 \\
	\hline
	Non-Overlapping Template & 0.507021 & 99/100 \\
	\hline
	Overlapping Template & 0.304126 & 99/100 \\
	\hline
	Approximate Entropy & 0.319084 & 99/100 \\
\hline
\hline
\end{tabular}
\end{centering}
\caption{NIST test suite results for our photon-generated random sequences.}
\label{tabla-nist}
\end{table}

\section{Conclusions}

Random number generators are usually assessed using a battery
of tests (such as the one from NIST \cite{art:nist}), which are quite practical but not based on a formal
definition of randomness. The Borel normality test employed here is based on algorithmic information
theory, which provides a mathematical framework which can formalize randomness.

In \cite{art:Calude:ExpEvi} it was reported that quantum random number sequences, 
built from photon detection events, failed in some cases to pass the Borel normality test,
while the pseudo-random sequences generated with computer codes had no problems
to fulfill the Borel normality requirements demanded by algorithmic randomness.

In this contribution we report an analysis of Borel normality of sequences of random numbers
generated from the time intervals between successive detection events in a photon-pair source based on  spontaneous parametric downconversion. They
pass comfortably the Borel normality test.

Before definite conclusions can be extracted from this comparison, we plan to carry out further experiments with longer photon-derived random
sequences.  We also
plan to carry out experiments with an attenuated laser instead of SPDC light, and with a beamsplitted introduced both in one arm of an SPDC 
source and on the path of an attenuated laser. We hope that these steps will help to clarify if, and under what circumstances, it is experimentally posible to distinguish random number sequences
created using quantum sources from computer-generated pseudo-random numbers. 

\section*{Acknowledgments}

This work was supported in part by grants from SEP-Conacyt(Mexico), PAPIIT (UNAM) IN-111212 and AFOSR
grant FA9550-13-1-0071.

%
%
%
%

\section*{References}
\bibliography{bibliografia.bib}

\end{document}